\documentclass[12pt]{article}
\usepackage{amscd,amsmath,amssymb}
\def\DH{\rm I\kern-1.5pt\rm H\kern-1.5pt\rm I}

\def\DR{\rm I\kern-1.45pt\rm R}
\def\DC{\kern2pt {\hbox{\sqi I}}\kern-4.2pt\rm C}

\newcommand{\ba}{\begin{array}}
\newcommand{\ea}{\end{array}}
\newcommand{\be}{\begin{equation}}
\newcommand{\ee}{\end{equation}}
\newcommand{\bea}{\begin{eqnarray}}
\newcommand{\eea}{\end{eqnarray}}
\newcommand{\bi}{\begin{itemize}}
\newcommand{\ei}{\end{itemize}}

\begin{document}
\thispagestyle{empty}
\begin{center}
{\bf \large Reductions related with  Hopf maps}\\
\vspace{0.5 cm} {\large  Vahagn Yeghikyan}
\end{center}

{\it Yerevan State University, 1 Alex Manoogian St.,   Yerevan,
0025, Armenia}
\begin{abstract}
We consider the reductions of $2p$-dimensional particle system
($p=2,4,8$), associated with the Hopf map. For the third Hopf map we
explicitly construct the functions associated to the symmetry
related to the rotations in the fiber.
\end{abstract}
\section{Introduction}

It is known that the systems describing motion of particle in the
field of  of Dirac an Yang monopoles can be constructed using the
reduction procedure associated with the first and second Hopf maps
\cite{gknty}. The Hopf maps(fibrations) are fibrations of spheres
over spheres with the fiber-sphere \cite{hopf1931}. There are four
Hopf maps: $S^{2n-1}/S^{n-1}=S^n$, ($n=1,2,4,8$). The Dirac and Yang
monopoles are related with  the first and the second ones
respectively. Zero Hopf map is related with anyons (or magnetic
vortices)\cite{ntt}, while  for the last Hopf map the mentioned
procedure does not exist.
 Moreover, it is even unclear which sort of monopole should arise after the reduction.
 The problem comes from the fact that  the algebra of the octonions is not associative
 (equivalently the fiber of the last fibration is not a group manifold). Therefore, the transformation,
 which leave invariant the coordinates of base of the third Hopf fibration are not isometries of bundle space.

The goal of current paper is to investigate the problems arising
while trying to construct the reduction procedure related to the
third Hopf map. For this purpose we formulate  the reduction
procedures associated with the  Hopf maps (\cite{gknty,smic}) in
terms of real coordinates and Clifford algebras.
% We show that for the first and second Hopf maps there is no need to add the Fayet-Iliopoulos-like
% term to the initial system\cite{Z,mic}:
% the full time-derivative term arises naturally within a consistent reduction procedure.

Together with geometric methods (for review, see e.g. \cite{lnp})
 the algebraic understanding of the nature of the Hopf maps leaves to no surprise that important differences
are encountered between the Hopf maps.

The paper is arranged as follows.

In the Second Section we present an explicit description of the  Hopf maps in terms  needed for our purposes.

In the Third section  we employ the Hopf maps
to reduce the bosonic free-particle systems to lower dimensional systems with magnetic $SU(2)$ monopoles.

\setcounter{equation}{0}
\section{Hopf maps}

The Hopf maps (or Hopf fibrations) are the fibrations of the
sphere over a sphere, $ {  S}^{2p-1}/{  S}^{p-1}={  S}^{p}$,
$p=1,2,4,8$. These fibrations reflect the existence of {\sl normed division algebras:} real
($\mathbb{R}$, $p=1$), complex ($\mathbb{C}$, $p=2$), quaternionic ($\mathbb{H}$, $p=4$) and octonionic
($\mathbb{O}$, $p=8$) numbers.
%
%It this section we consider only the first and second Hopf maps: \be
% {  S}^3/{  S}^1={  S}^2\quad ({\rm first}\;{\rm Hopf}\;{\rm map}), \qquadgenerating
%{  S}^7/{  S}^3={  S}^4\quad ({\rm second}\;{\rm Hopf}\;{\rm map}).
%  \label{hm0}
% \ee
% We omit consideration of the zero Hopf map because its triviality,
% and of the third Hopf map because its difficulty.
\subsection{Normed Division algebras}

Any element of normed division algebras can be expressed via the generating elements of the algebra ${\bf e}_\mu$.
\be
{\bf x}=x^n+x^\mu{\bf e}_\mu,\quad \mu=1,\ldots,n-1,\quad n=1,2,4,8,\label{elems}
\ee
where the generating elements satisfy the following  multiplicative rule:
\be
{\bf e}_\mu{\bf e}_\nu=-\delta_{\mu\nu}+C_{\mu\nu\lambda}{\bf e}_\lambda,\label{genmult}
\ee
where $C_{\mu\nu\lambda}$ are constants antisymmetric under any permutations of indices.
 Here and in the further we will use bold style to denote the elements of algebra and normal style for real elements.

The conjugation and norm are defined by analogy with complex numbers($n=2$):
$$
{\bf \bar{x}}=x_n-x_\mu{\bf e}_\mu,\quad |{\bf x}|\equiv\sqrt{{\bf x}{\bf \bar{x}}}=\sqrt{x_ax_a}.
$$
The greek symbols $\mu,\nu,\lambda$ run $1,\ldots,n-1$, while the latin symbols $a,b,c=1,\ldots,n$.

It was proven\cite{ndae}, that one can construct the constants $C_{\mu\nu\lambda}$ so that the algebras have division operation only for the dimensions $n=1,2,4,8$. It is clear that for real and complex numbers we have $C_{\mu\nu\lambda}=0$. For quaternionic numbers we define $C_{\mu\nu\lambda}=\varepsilon_{\mu\nu\lambda}$, where $\varepsilon_{\mu\nu\lambda}$ are the elements of totally antisymmetric tensor and for the octonions we have
\be
C_{123}=C_{147}=C_{165}=C_{246}=C_{257}=C_{354}=C_{367}=1,
\ee
while all other non-vanishing components are determined by the total antisymmetry.

Each time we pass from a lower dimensional algebra to the next one,
we lose some symmetry. Hence, first we lose the fact that every
element is its own conjugate, then we lose commutativity, then we
lose associativity \cite{baez}. However, the last- octonionic
algebra has a weaker property called {\sl alternativity} which
implies any subalgebra consisting of two elements is associative(for
associativity we should have three). For more information about
modern status of the theory of normed division algebras see an
excellent review \cite{baez}.

One can consider the elements  \eqref{elems} as columns with $n$ real elements. Using \eqref{genmult} one can write down the multiplicative rule for this columns:
\be
(xy)_a=x_c \gamma^c_{ab}y^b,\quad (\gamma^c)_{ab}=-\delta_{an}\delta^c_b+\delta^c_a\delta_{bn}+\delta^c_n\delta_{ab}-C^{\mu}_{ab}\equiv(\gamma^c)_{ab},\label{gammadef}
\ee
where we have chosen $C_{abc}=0$ if at least one index is equal to $n$.
Since we deal with Euclidean space, there is no difference between upper and lower indices. Here and further we will denote the columns by the normal letters without indices.

One can see, that from requirement $|{\bf x}{\bf y}|=|{\bf x}||{\bf y}|$ ($\forall {\bf x},{\bf y}\in \mathbb{R},\mathbb{C},\mathbb{H},\mathbb{O}$) and the definition \eqref{gammadef} it follows, that
\be
\left\{\gamma^\mu,\gamma^\nu\right\}=-\delta^{\mu\nu},\quad \gamma^n={\bf 1}_{n\times n},\quad (\gamma^\mu)^T=-\gamma^\mu,\label{gammarel}
\ee
where $\left\{.,.\right\}$ denotes anticommutator and $T$- transpose of matrices(See e.g. \cite{gunket}).
This is all we need to know about the normed division algebras. Now, let us pass to the description of Hopf maps.

\subsection{Hopf maps (Normed division algebras)}
Let us describe the  Hopf maps using normed division algebras.
For this purpose, we consider the functions ${\bf x}(u_\alpha,{\bar
u}_\alpha), x_{p+1}(u_\alpha,{\bar u}_\alpha)$

\be {\bf x}=2{\bf \bar u}_1{\bf u}_2 ,\quad x^{n+1}={\bf \bar
u}_1{\bf u}_1-{\bf \bar u}_2{\bf u}_2, \label{hm}\ee
where ${\bf
u}_1,{\bf u}_2$ are complex numbers for $n=2$ case (first Hopf map),
 quaternionic numbers for the $n=4$ case (second Hopf map)  and octnionic numbers for $n=8$ (third Hopf map)
 (see,e.g. \cite{lnp}). One
can consider them as coordinates of the $2n$-dimensional  space
$\mathbb{R}^{2n}$ ($n=2$ for ${\bf u}_{1,2}$ complex numbers; $n=4$ for
${\bf u}_{1,2}$ quaternionic numbers, $n=8 $ for ${\bf u}_{1,2}$ octonionic numbers).
In all  cases  $x_{n+1}$ is a
real number while ${\bf x}$ is, respectively,  complex number
($n=2$),  a quaternionic one ($n=4$), and octonionic one ($n=8$)
\be {\bf x}\equiv x^{n}
+{\bf e}_\mu x^\mu.\ee
One could immediately
check that the following equation holds: \be r^2\equiv {\bf \bar
x}{\bf x}+(x^{n+1})^2=( {\bf\bar u}_{1}{\bf u}_1+{\bf \bar u}_2{\bf
u}_2)^2\equiv R^4. \label{hm1}\ee
 Thus, defining the $(2n-1)$-dimensional sphere in
$\\mathbb{R}^{2n}$ of radius $R$, ${\bf\bar u}_\alpha{ \bf u}_\alpha =R^2$,
we will get the $p$-dimensional sphere in $\mathbb{R}^{n+1}$ with radius
$r=R^2$.

The expressions (\ref{hm}) can be easily inverted by the use of equality (\ref{hm1}) \be
{\bf u}_\alpha={\bf g} r_\alpha ,\label{inve}\ee  where $$
\quad r_1=\sqrt{\frac{r+x^{n+1}}{2}},\quad  r_2\equiv
r_+=\frac{{\bf x}}{\sqrt{2(r+x^{n+1})}}
,\quad {\bar {\bf g}}{\bf g}=1 .$$

%It is seen that \be
% {\bf g}^2={\bf u}_2({{\bf\overline u}}_2)^{-1}.
%\label{ks0}\ee
It follows from the last equation in (\ref{inve}) that ${\bf g}$
parameterizes the $(n-1)$-dimensional sphere of unit radius.\\

Using  above equations, it is easy to describe the first three Hopf maps. Indeed, for
$n=1,2,4$
the functions  ${\bf x}, x_{n+1}$ remain invariant under the
 transformations
\be {\bf u}_\alpha\to {\bf{G}} {\bf u}_\alpha ,\quad {\rm
where}\quad {\bar {\bf {G}}}{\bf{G}}=1 %\Rightarrow \left\{
%\begin{array}{lll}
%{\bf {G}}=\lambda & |\lambda_1|^2=1 \Leftrightarrow \lambda=\pm
%1&{\rm for}\quad p=1 \cr {\bf {G}}=\lambda_1+{\bf i}\lambda_2 &
%|\lambda_1|^2+|\lambda_2|^2=1 &{\rm for}\quad p=2 \cr {\bf
%G}=\lambda_1+{\bf i}\lambda_2+{\bf j}\lambda_3+{\bf k}\lambda_4 &
%|\lambda_1|^2+\ldots+|\lambda_4|^2=1&{\rm for}\quad p=4 .
%\end{array}\right.
\label{G}\ee Therefore,  ${\bf G}$ parameterizes the spheres $S^{n-1}$ of
unit radius. Taking into account the  isomorphism
 between these spheres and the groups,for $n=1,2,4$:$S^0=\mathbb{Z}_2$, ${ S}^1=U(1)$, ${  S}^3=SU(2)$,
we get that (\ref{hm})  is invariant under $G-$group
transformations for $n=1,2,4$ (where $G=\mathbb{Z}_2$ for $n=1$,  $G=U(1)$ for $n=2$, and $G=SU(2)$ for
$n=4$).\\

For the octonionic case $n=8$ situation is more complicated.
Because of losing associativity the standard
transformation ${\bf u}_\alpha$ that leaves invariant
coordinates ${\bf x}, x_9$ will not be just (\ref{G}).
Instead, we should write

Its modification can be easily obtained using
(\ref{inve}): \be {\bf u}_\alpha \mapsto ({\bf G}{\bf g})({\bf
\bar g}{\bf u}_\alpha)= \frac{({\bf G}{\bf u}_1)({\bf
\bar u}_1{\bf u}_\alpha)}{{\bar{\bf u}_1 {\bf u}_1}} . \label{octtrans} \ee
Also, for the $n=8$ case the bundle $S^7$ is not isomorphic  with any group. So, one can expect further troubles in the extensions
 of the constructions, related with the lower Hopf maps to the third one.

\subsection{Hopf maps (Spinor representation)}

One can consider a $2n$-dimensional column $U$ consisting of the real coordinates $u_{\alpha,a}$, $\alpha=1,2$, $a=1,\ldots,n$:
$$
U=(u_{1,1},\ldots,u_{1,n},u_{2,1},\ldots,u_{2,n}).
$$
Using this denotation one can rewrite \eqref{hm} in the following form:
\be
x^A=U\Gamma^A U,\quad A=1,\ldots, n+1
\ee
where
\be
\Gamma^\mu=\left(\begin{array}{cc}
                0 & \lambda^\mu\\
                -\lambda^\mu & 0
               \end{array}
\right),\quad \Gamma^{n}=\left(\begin{array}{cc}
                0 & {\bf 1}_{n\times n}\\
                {\bf 1}_{n\times n} & 0
               \end{array}
\right),\quad \Gamma^{n+1}=\left(\begin{array}{cc}
                -{\bf 1}_{n\times n} & 0\\
                0 & {\bf 1}_{n\times n}
               \end{array}
\right),
\ee
where
\be
(\lambda^\mu)_{ab}=-\delta_{an}\delta^\mu_b+\delta^\mu_a\delta_{bn}+C_{\mu ab}\label{lambdadef}
\ee
(compare with \eqref{gammadef}). This matrices satisfy the  relations  \eqref{gammarel}. One can check, that the matrices $\Gamma^A$ are the Euclidean gamma-matrices satisfying the anticommutational relations:
\be
\left\{\Gamma^A,\Gamma^B\right\}=\delta^{AB}{\bf 1}_{2n\times 2n}.
\ee

In \cite{mny} all the three Hopf maps were explicitly constructed using spinor representations of $SO(1,n+1)$. For the complex and quaternionic ($n=2,4$) case it was shown the direct connection between this description and the one using normed division algebras.

It is obvious, that for $n=2,4$ we have reducible representation of Clifford algebra(we have $n=2,4$ generating elements and 4 and 8 dimensional representation respectively).

$n=8$ is the only case for which the constructed matrices form an irreducible representation of Clifford algebra. E.g. nine matrices $\Gamma^A$ have dimension $16=2^{[9/2]}$. It is clear, that in this representation all the matrices $\Gamma^A$ are symmetric. These is, in fact, such representation, where the matrix of charge conjugation is identity matrix:
$$
({C^T})^{-1}\Gamma^A C=(\Gamma^A)^T =\Gamma^A,\quad C={\bf 1}_{16\times16}
$$

% (To be commented)
% The transformation matrix $T$ from this representation to the chiral one looks as follows:
% \be T=
% \left(
% \begin{array}{cccccccccccccccc}
%  1 & -1 & -i & i & i & i & 1 & 1 & 0 & 0 & 0 & 0 & 0 & 0 & 0 & 0 \\
%  i & i & 1 & 1 & -1 & 1 & i & -i & 0 & 0 & 0 & 0 & 0 & 0 & 0 & 0 \\
%  1 & 1 & -i & -i & -i & i & -1 & 1 & 0 & 0 & 0 & 0 & 0 & 0 & 0 & 0 \\
%  -i & -i & 1 & 1 & 1 & -1 & i & -i & 0 & 0 & 0 & 0 & 0 & 0 & 0 & 0 \\
%  1 & -1 & i & -i & i & i & -1 & -1 & 0 & 0 & 0 & 0 & 0 & 0 & 0 & 0 \\
%  i & -i & -1 & 1 & 1 & 1 & i & i & 0 & 0 & 0 & 0 & 0 & 0 & 0 & 0 \\
%  -1 & -1 & -i & -i & i & -i & -1 & 1 & 0 & 0 & 0 & 0 & 0 & 0 & 0 & 0 \\
%  -i & i & -1 & 1 & -1 & -1 & i & i & 0 & 0 & 0 & 0 & 0 & 0 & 0 & 0 \\
%  0 & 0 & 0 & 0 & 0 & 0 & 0 & 0 & 1 & -1 & -i & i & -i & -i & -1 & -1 \\
%  0 & 0 & 0 & 0 & 0 & 0 & 0 & 0 & i & i & 1 & 1 & 1 & -1 & -i & i \\
%  0 & 0 & 0 & 0 & 0 & 0 & 0 & 0 & 1 & 1 & -i & -i & i & -i & 1 & -1 \\
%  0 & 0 & 0 & 0 & 0 & 0 & 0 & 0 & -i & -i & 1 & 1 & -1 & 1 & -i & i \\
%  0 & 0 & 0 & 0 & 0 & 0 & 0 & 0 & 1 & -1 & i & -i & -i & -i & 1 & 1 \\
%  0 & 0 & 0 & 0 & 0 & 0 & 0 & 0 & i & -i & -1 & 1 & -1 & -1 & -i & -i \\
%  0 & 0 & 0 & 0 & 0 & 0 & 0 & 0 & -1 & -1 & -i & -i & -i & i & 1 & -1 \\
%  0 & 0 & 0 & 0 & 0 & 0 & 0 & 0 & -i & i & -1 & 1 & 1 & 1 & -i & -i
% \end{array}
% \right)
% \ee
% Hence, the chiral representation $\Gamma_{ch}^A$ can be obtained:
% \be
% \Gamma_{ch}^A=T^{-1}\Gamma^A T
% \ee
% (/To be commented)

In \cite{mny} it was shown, that the infinitesimal transformation \eqref{octtrans} can be presented in the following form:
\be
\delta U=-\frac{1}{6}\omega_{AB}(U^T\Gamma^{ABCD}U)\Gamma^{CD}U
\ee
\section{Hopf maps and reductions}

Let us apply the obtained formulae for the Hopf fibrations to reduce the $2n$-dimensional free particle system to a lower dimensional one. For this reason we consider Lagrangian in terms of the coordinates of fiber and base of Hopf fibrations:
\be
 {\cal L}_{2n}=
 \frac{g(r_\alpha)}{2}\left(\dot{\bar{\bf r}}_\alpha\dot{{\bf r}}_\alpha+
 2Re\left((\dot{\bar{\bf r}}_\alpha\bar{\bf g})(\dot{\bf g}
 {\bf r}_\alpha)\right)+r\dot{\bar{\bf g}}\dot{\bf g}\right)
 =\label{lag0}
\ee
$$
=\frac{{g}}{2}\left( \dot{{\bf r}}_\alpha{\bf \dot{\bar{r}}}_\alpha+
 2r v_a A_{ab}\dot{v}_b+r  \dot{v}_a\dot{v}_a\right),\qquad a,b,c,d=1,\ldots,8
$$
 where \be A_{ab}=\frac
{x_c(\Sigma^{cd})_{ab}\dot{x}_d}{2r(r+x_{9})},\quad \Sigma^{\mu\nu}=
\frac{\left[\lambda^\mu,\lambda^\nu\right]}{2},\quad\Sigma^{\mu
n}=-\Sigma^{n\mu}=\lambda^\nu, \qquad \mu,\nu=1,\ldots,7\ee
is precisely
the potential of $U(1)$ Dirac, $SU(2)$ Yang  and $S0(8)$ monopoles \cite{tchrakian} with $\lambda_i$
be the two, four or eight-dimensional gamma-matrices (for $n=2,4,8$ respectively). The functions $v_i$ are
the Euclidean coordinates of the fiber of Hopf fibrations: $S^1$ for the first, $S^3$ for the second and $S^7$ for the third Hopf map. ${\bf
g}=v_8+{\bf e}_av_a$, $v_av_a=1$ and express via projective coordinates of $S^{2n-1}$ as follows:
\be v_n=\frac{1-y^2}{1+y^2},\quad v_\mu=\frac{2y_\mu}{1+y^2},\quad
y^2=\sum\limits_{\mu=1}^{n-1}y_\mu^2 \ee

Taking into account the last expressions, we can represent
the  Lagrangian in the following form:
$$
 {\cal L}=\frac{{g}}{2}\left({\bf \dot{\bar{ r}}}_\alpha{{\bf\dot r}}_\alpha+2rD_\mu\dot{y}_\mu+
 2r\frac{\dot{y}_\mu\dot{y}_\mu}{(1+y^2)^2}\right),$$
 where\be
 D_\mu=\frac{1}{(1+y^2)^2}\left(A_{n\mu}(1-y^2)+y_\nu A_{\nu\mu}+2y_\nu A_{n\nu}y_\mu\right),\quad n=2,4,8.
\ee

Let us replace our Lagrangian by the variationally
 equivalent one, performing Legendre transformation of the ``isospin" varyables $y_\mu$.

After some work we find
\be
{\cal L}_{int}=p_\mu\dot{y}_\mu+
\frac{g}{2}{\bf \dot{\bar{ r}}}_\alpha\dot{{\bf r}}_\alpha-(1+y^2)^2\frac{(p_\mu-2grD_\mu)^2}{8rg}
\ee
The generator of transformations (\ref{octtrans}) is defined, in these terms, as follows
\be
I^\mu=\frac{1}{2}(1+y^2)S^\mu_\nu\frac{p_\nu}{2},\quad{\rm where}\quad
S^\mu_\nu=\frac{1}{1+y^2}\left(2y_\mu y_\nu+(1-y^2)\delta^\mu_\nu+2 y_\lambda C^\mu_{\nu \lambda}\right),
\ee
with $C_{\mu\nu\lambda}$ be structure constants of complex, quaternionic and octonionic algebra.

{\bf Remark.} It is obvious, that for the case of associative algebras(complex and quaternionic) the transformation \eqref{octtrans} form the symmetry of the initial Lagrangian \eqref{lag0}. However, for the case of octonions the lack of associativity leads to the fact, that this transformations do not preserve the Lagrangian and, therefore, the quantities $I_\mu$ are not the integrals of motion of the system. We will discuss this below.

Taking into account the equalities
\be
SS^T={\bf 1}_{n-1},\qquad
 \frac{I^\mu I^\mu}{2{g}r}=(1+y^2)^2\frac{p^2}{16r{g}}\qquad\dot{r}_A\dot{r}_A-rD_\mu D_\mu(1+y^2)^2=
 \frac{\dot{x}_A\dot{x}_A}{4r}
\label{i2}\ee
we can represent the Lagrangian in very transparent form
\be
{\cal L}_{int}={g}\frac{\dot{x}_A\dot{x}_A}{8r}+p_\mu\dot{y}_\mu+r{g}J_{ab}A_{ab}-\frac{1}{4}\frac{I_\mu I_\mu}{2gr},\label{unfin}
\ee
where
\be
J_{\mu\nu}=y_\mu p_\nu-y_\nu p_\mu,\quad J_{\mu n}=-J_{n\mu}=\frac{1-y^2}{2}p_\mu+( y_\nu p_\nu)y_\mu, \quad n=2,4,8.
\ee
are the generators of $SO(n)$ rotations.

\subsection{$n=2$ complex case}
In this case we have $a,b=1,2$ and, therefore, one element $J_{ab}$ : $J_{12}=-J_{21}=p$. It is easy to check, that this element is a constant of motion of the system and therefore we can fix its value to be equal to a constant $s$. The term with $\dot{y}$ disappears because it becomes full time derivative and finally we find the reduced Lagrangian:
\be
{\cal L}_{3}={g}\frac{\dot{x}_A\dot{x}_A}{8r}+r s A_{D}-\frac{s^2}{2gr},\quad A=1,2,3,
\ee
where $A_D$ is the vector-potential of Dirac monopole.
\subsection{$n=4$ quatrenionic case}

We have already mentioned, that for $n=4$ the representation of Clifford algebra is not minimal and, therefore not all the components of $A_{ab}$ are independent. Using the properties of $\varepsilon_{\mu\nu\lambda}$ one can find the following connection between this elements:
\be
\varepsilon_{\lambda\mu\nu}A_{\mu\nu}=2A_{\lambda n}
\ee
And, therefore, we find
\be
J_{ab}A_{ab}=P_\mu \tilde{A}_\mu
\ee
where
\be
\tilde{A}_\lambda=\frac{1}{2}\varepsilon_{\lambda\mu\nu}A_{\mu\nu},\quad P_\lambda=J_{n\lambda}-\frac{1}{2}\varepsilon_{\lambda\mu\nu}J_{\mu\nu}.
\ee
Let us mention that the following identity obeys:
\be
I_\mu=-J_{n\lambda}-\frac{1}{2}\varepsilon_{\lambda\mu\nu}J_{\mu\nu}
\ee
Using this denotations, one can rewrite the Lagrangian \eqref{unfin} as follows:
\be
L_{8}={g}\frac{\dot{x}_A\dot{x}_A}{8r}+p_\mu\dot{y}_\mu-4r{g}P_\mu \tilde{A}_{\mu}-\frac{1}{4}\frac{I_\mu I_\mu}{2gr}\label{almostreduced}
\ee
The quantities $P_\mu$ together with $I_\mu$ form $so(4)=so(3)\times so(3)$ algebra of symmetries of $S^3$. E.g. they obey the following commutation relations:
\be
\left\{P_\mu,I_\nu\right\}=0,\quad \left\{P_\mu,P_\nu\right\}=\varepsilon_{\mu\nu\lambda}P_\lambda,\quad \left\{I_\mu,I_\nu\right\}=\varepsilon_{\mu\nu\lambda}I_\lambda,\quad I_\mu I_\mu=P_\mu P_\mu\label{so4}
\ee

Now, we are ready to fix the values of the integrals of motion and hence, to perform the reduction. Without loss of generality we can fix
\be I_1=I_2=0,\quad I_3=s\ee
Because of the relations \eqref{so4} we can denote:
\be
P_+=P_2+\imath P_1=-\imath s\frac{\bar z}{1+z \bar z}\equiv- \imath s h_-,\quad P_-=\bar P_+=\imath s\frac{z}{1+z \bar z}\equiv \imath s h_+\label{hdef}\ee
$$
 P_3=-s\frac{1-z\bar z}{1+z \bar z}\equiv -sh_3,\quad \left\{z,\bar z\right\}=(1+z\bar z)^2
$$
and the Lagrangian \eqref{almostreduced} will take the following form:
\be{\cal
L}_{red}=\frac{\widetilde{g}\dot{x}_{A}\dot{x}_{A}}{2}- {\bf i}
s\frac{\bar {z}\dot{z}-z \dot{\bar z}}{1+z\bar z} -{sh_\mu(z,\bar
z)}A_\mu-\frac{s^2}{2r^2\widetilde{g}},\qquad
{\widetilde{g}}\equiv\frac{g}{2r}, \qquad \mu=1,\ldots,5, \ee
where the quantities $h_\pm,h_3$ are defined by \eqref{hdef}.

The second term in the above
reduced Hamiltonian is the one-form defining the symplectic (and
K\"ahler) structure on ${  S}^2$, while $h_k$ given in (\ref{hdef}) are the Killing potentials defining the isometries of the
K\"ahler structure. We have in this way obtained the Lagrangian describing the motion of a five-dimensional isospin particle
in the field of an $SU(2)$ Yang monopole. The metric of the configuration space is defined
by the expressions ${\widetilde g}_{\mu\nu}=
\frac{g}{2r}\delta_{\mu\nu}$. For a detailed description of
the dynamics of the isospin particle we refer to  \cite{horv}.

\subsection{$n=8$ octonionic case}

It was already mentioned that because of lack of associativity of the octonionic algebra the transformations \eqref{octtrans} and therefore the functions $I_\mu$ do not form isometries of the Lagrangian \eqref{lag0}.

It is seen from (\ref{i2}), that $I_aI_a=J_{ij}J_{ij}$ defines constant of motion of the system, in complete
analogy with the lower Hopf map. Reducing the system by this constant of motion, we shall get the system with
$30 (=2\cdot9+12)$-dimensional phase space, which describes the interaction of the $8$-dimensional isospin particle
with $S0(8)$ monopole field. The dimensionality of the internal phase space of the particle is equal to $12$.

Now, we should proceed the last step: we need to modify the
Lagrangian by adding the specific term (vanishing for the lower Hopf
maps), in such a way, that not only $I^2$, but each $I_a$ will be
the constant of motion of our system.

\section{Conclusion}
We have presented the reduction procedure associated with the first and second Hopf map in the Lagrangian approach. For the last- the third Hopf fibration we have presented the explicit formulae of the Lagrangian in coordinates of base and fiber. Since we deal with irreducible representation of $SO(8)$ algebra it is impossible to construct the motion integrals corresponding to respective ones for the first and second Hopf maps. The only way to avoid this problem seems to be modifying the initial Lagrangian or considering non- Lie algebras of motion integrals.

\par~\par
{\bf {\large Acknowledgments.}}
\par~\par
I am grateful to Armen Nersessian, Francesco Toppan, Zhanna
Kuznetsova and Marcelo Gonzales for collaboration  on the work
\cite{gknty} which became the base for the current one.
Special thanks to Armen Nersessian for the useful discussions and
for the help in preparation of this paper.  I would like to thank
George Pogosyan for given opportunity to give a talk at  XIV
International Conference on {\sl Symmetry Methods in Physics}(16-22
August, 2010), which hold In Tsakhkadzor, Armenia.

 The work was partially
supported by Volkswagen Foundation I/84 496 grants.

\end{document}